%% file: Potts_APSOS_proof.tex
\documentclass[aps,pre,10pt,twocolumn,letterpaper,amsmath,amssymb]{revtex4-2}

\usepackage[utf8]{inputenc}
\usepackage{graphicx}
\usepackage{subfigure}
\usepackage[usenames,dvipsnames]{xcolor}
\usepackage{epstopdf}
\usepackage{amsmath,amsthm,amssymb}
\usepackage{tcolorbox}
\usepackage{latexsym}
\usepackage{bm}
\usepackage{dcolumn}
\usepackage{braket}
\usepackage[normalem]{ulem}
\usepackage{times}
\usepackage{hyperref}
\usepackage{enumitem}

\newcommand{\ignore}[1]{}
\newcommand{\red}[1]{#1}

\newcommand{\resub}[1]{#1}
\newcommand{\blue}[1]{#1}

\newcommand{\qedblack}{\hfill $\blacksquare$}

\newcommand{\typea}{type-\emph{a}}
\newcommand{\typeb}{type-\emph{b}}

\begin{document}


\title{Half-ice, half-fire-driven ultranarrow phase crossover in one-dimensional\\
decorated $q$-state Potts ferrimagnets: An AI-co‑led exploration}
\author{Weiguo Yin}
\email{wyin@bnl.gov}
\affiliation{Condensed Matter Physics and Materials Science Division,
Brookhaven National Laboratory, Upton, New York 11973, USA}

\begin{abstract}
OpenAI's reasoning model \texttt{o3-mini-high} was used to carry out an exact analytic study of one-dimensional ferrimagnetic site- and bond-decorated $q$-state Potts models. We demonstrate that the finite-temperature ultranarrow phase crossover (UNPC), driven by a hidden ``half-ice, half-fire'' state recently discovered in the $q=2$ case (Ising model), persists for $q>2$. \resub{Moreover,} we identify unique features for $q>2$, including the dome structure in the field-temperature phase diagram, and for large $q$ a secondary high-temperature UNPC to the fully disordered paramagnetic state. 
\blue{As the UNPC quickly approaches a genuine transition by enhancing $J$, the interaction between the backbone spins, two distinct behaviors emerge: In the site-decorated Potts model, $T_0$ is independent of $J$ and thus remains unchanged (Type-I UNPC), and in the bond-decorated Potts model with $q > 2$, $T_0$ depends on $J$ and quickly shifts toward a finite temperature as $J$ increases (Type-II UNPC).}  These results establish a versatile framework for engineering controlled fast state-flipping switches in low-dimensional systems. 
Our nine-\red{dan} artificial intelligence (AI)-contribution 
\red{framework} assigns AI the meritorious status of AI-co-led discovery in this work.\\

\noindent DOI: \href{https://doi.org/10.1103/425p-bt1x}{10.1103/425p-bt1x}

\end{abstract}

{
}
\date{Received 4 December 2025; accepted 8 June 2026; published 22 June 2026}

\maketitle


\section{Introduction}

The one-dimensional (1D) Ising\blue{, Potts,} and Heisenberg models with short-range interactions are textbook models in statistical mechanics and the theory of magnetism, but are prohibited from exhibiting finite-temperature phase transitions~\cite{Kivelson_24_book_statistical,Mattis_book_08_SMMS,Mattis_book_1981,Mattis_book_1985,Baxter_book_Ising,Ising1925,Mermin_PRL_theorem,Cuesta_1D_PT,Potts_1952,Potts_RMP_82}. As a result, they have been largely overlooked for their potential in both fundamental research and technological applications. Recently, 
\resub{the 1D Ising model attracted renewed attention because} \blue{geometric frustration induced by various spin decorations was found to drive phase-transition-like thermodynamic behavior at finite temperature $T_0$ with a narrow transition width $2\delta T$, a phenomenon termed pseudotransition}~\resub{\cite{005_Galisova_PRE_15_double-tetrahedral-chain,007_Torrico_PRA_16_Ising-XYZ-diamond-chain,009_review_Souza_SSC_18_Ising-XYZ-diamond-chain_double-tetrahedral-chain-spin-electron,010_Carvalho_JMMM_18_Ising-XYZ-diamond-chain_quantum-entanglement,Carvalho_AnnalsPhys_19_Ising-XYZ,013_Rojas_PRE_19_previous_4_models,014_Rojas_JPC_20_Ising-Heisenberg_spin-1-double-tetrahedral-chain,011_Rojas_BJP_20_Ising-Heisenberg-tetrahedral_diamond,015_Strecka_APPA_20_Ising-diamond-chain,016_Strecka_book_chapter,017_Krokhmalskii_PA_21_3-previous-chains_effective_model,Rojas_PRE_26}} \blue{whose underlying mechanism follows directly from the ground-state phase diagram: at zero temperature, place the system near a phase boundary between two phases with macroscopically different degeneracy, on the side with less entropy; at finite $T_0$, thermal fluctuations drive the system into the other phase, which has lower free energy due to its higher entropy~\cite{Miyashita_10_review_frustration}. The pseudotransition width $2\delta T$ can be reduced by placing the system closer and closer to the phase boundary, since $\delta T=0$ corresponds to a genuine phase transition; however, since genuine phase transitions in the 1D Ising model---despite decorations---occur only at zero temperature~\cite{Cuesta_1D_PT}, $T_0\to 0$ inevitably as $2\delta T \to 0$ for pseudotransitions~\cite{017_Krokhmalskii_PA_21_3-previous-chains_effective_model}\resub{, posing a fundamental limitation for applications}.} 

\resub{Thus,} \blue{the concept of ultranarrow phase crossover (UNPC) was introduced, defined by the independence of the crossover temperature $T_0$ and width $2\delta T$ so that $2\delta T$ can be reduced exponentially by tuning the interactions, while $T_0$
remains fixed at a much higher temperature, approaching a forbidden finite-temperature phase transition arbitrarily closely,} \resub{circumventing the century-old prohibition in the 1D Ising model}\blue{~\cite{Yin_MPT,Yin_MPT_chain}. It has been demonstrated that UNPC can spontaneously emerge in 1D Ising ladders with infinitely many ways of decorating the rungs~\cite{Yin_MPT} and that UNPC can occur in bond-decorated single chains in the presence of an external magnetic field~\cite{Yin_MPT_chain}. Moreover, UNPC was shown to exist in 1D Ising ferrimagnets even without conventional geometric frustration, arising from an unexpected mechanism---exploiting a macroscopically degenerate excited state that is hidden in the ground-state phase diagram---whose significance was only fully revealed upon the discovery of the twin relationship between the half-fire, half-ice state~\cite{Yin_g} and the exotic half-ice, half-fire state (where half of the spins are fully frozen, while the remainder are completely disordered)~\cite{Yin_Ising_III_PRL}, with the latter driving the UNPC. This ice-fire spin reversal mechanism was further established across site-decorated one-dimensional, two-dimensional, and three-dimensional Ising models~\cite{Yin_site}.} It is thus natural to ask whether UNPC and the ice-fire mechanism extend to other domains of statistical models, thereby providing a general materials design guide~\cite{Ramirez_25_SmMn2Ge2}.

The $q$-state Potts model is a generalization of the Ising model ($q=2$) and serves as a bridge to study the crossover from discrete (Ising) to continuous (Heisenberg) symmetry~\cite{Potts_1952,Potts_RMP_82}. Examples \resub{range from $q=3$ for the out-of-plane stacking of charge stripes in La$_{1.67}$Sr$_{0.33}$NiO$_4$~\cite{ABC-AB_La2SrNiO4_PRL_13}} to $q=21$ for protein sequences~\cite{Potts_protein_PRE_13}. 
\resub{Yet, the Potts model is known by its own name---not as a generalized Ising model---because it poses significant number-theory-like challenges and NP problems in statistical mechanics and  computer science~\cite{Baxter_book_Ising}.} \blue{For example, the exact solutions of 1D $q$-state Potts models can be generally formulated in terms of the transfer matrix $\mathbb{T}$, but the order of $\mathbb{T}$ is $q^L$ (where $L$ is the number of spins per unit cell)~\cite{Kramers_Wannier_PR_41_transferMatrix,Glumac_JPA_93_Potts_LR}---Even the basic 1D $J_1$-$J_2$ Potts model with nearest-neighbor interaction $J_1$ and next-nearest-neighbor interaction $J_2$, where the order of $\mathbb{T}$ is $q^2$, had been solved in closed form for $q=2$ only} \resub{in the 1960s~\cite{Dobson_JMathP_69_Many-Neighbored-Ising-Chain}, until March 2025 for general $q$~\cite{Yin_Potts_J1-J2_1D}.} \resub{It is thus interesting to explore the much richer phase behavior offered by $q=3,4,5,\ldots,\infty$.} Recently, the \resub{prototypical} pseudotransition in the Ising model \resub{on a diamond chain~\cite{016_Strecka_book_chapter,017_Krokhmalskii_PA_21_3-previous-chains_effective_model}} was found to persist in the corresponding Potts model with  $q>2$~\cite{Panov_PRE_23_Potts,Panov_PRE_21_Potts}. 

\begin{figure}[tb]
    \begin{center}
\includegraphics[width=\columnwidth,clip=true,angle=0]{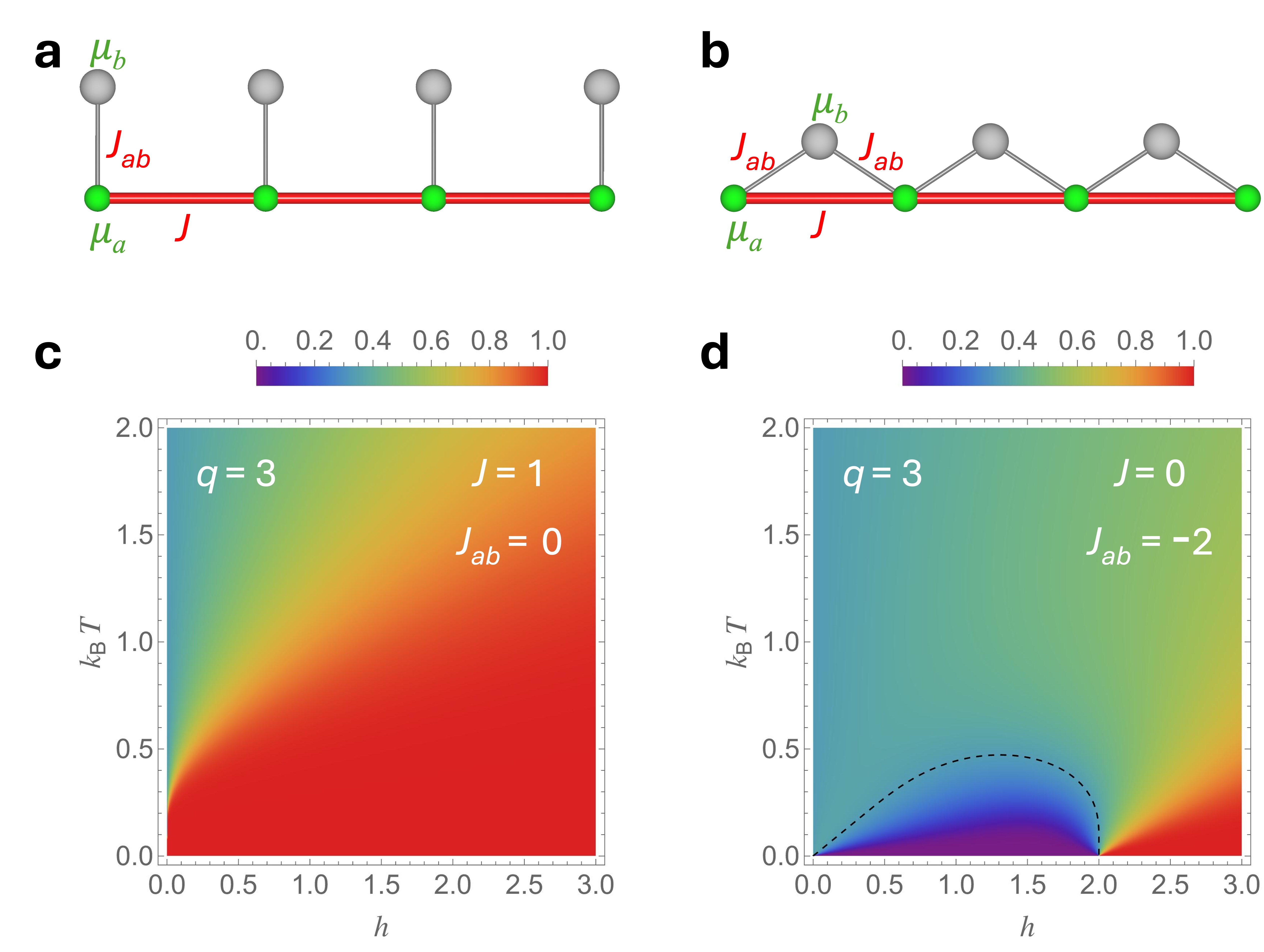}
    \end{center}
    \vspace{-0.5cm}
\caption{Schematics of the 1D decorated $q$-state Potts model with (a) site decoration and (b) bond decoration. The green and gray balls depict the {\typea} and {\typeb} spins with magnetic moments $\mu_a$ and $\mu_b$, respectively. The red bonds depict the  ferromagnetic interaction $J>0$ between {\typea} spins. The gray bonds depict the antiferromagnetic coupling $J_{ab}<0$ between  {\typea} and {\typeb} spins. Density plots of the backbone magnetization $m_a=\langle \delta(\sigma_i,1) \rangle$ as functions of the magnetic field $h$ and temperature $T$ in the common three-state site-decorated Potts models with (c) $J=1, J_{ab}=0$ and (d) $J=0, J_{ab}=-2$ show no indication of a UNPC. The dashed line in panel (d) is the contour line of $m_a=1/q$, which will quickly become a phase boundary line as $J$ increases; cf., Fig.~\ref{fig:magnetizations} for $q=3$. \resub{The color scale maps $m_a$ from 0 (purple) to 1 (red).}}
\label{Fig:structure}
\end{figure}

The purpose of this paper is to demonstrate that the half-ice, half-fire-driven UNPC not only persists in 1D decorated $q$-state Potts models 
for $q>2$, but also exhibits distinct features such as a $T_0$ dome in the $h$-$T$ (magnetic field versus temperature) phase diagram and for large $q$ a high-temperature UNPC to the fully disordered paramagnetic state. We present the studies of the standard 1D $q$-state Potts models with minimal site decoration [Fig.~\ref{Fig:structure}(a)] and bond decoration [Fig.~\ref{Fig:structure}(b)] in Secs.~\ref{section:site} and \ref{section:bond}, respectively. The results for the two kinds of decorated models are similar in the parameter space of interest---with one significant difference: As the interaction between the backbone spins increases, $T_0$ remains unchanged in \resub{both site- and bond-decorated models for $q=2$ and only in} the site-decorated model for $q>2$ (referred to as Type-I UNPC from now on); in contrast, $T_0$ shifts toward a fixed finite temperature in the bond-decorated Potts model with $q>2$, giving rise to Type-II UNPC. The collective nature of UNPC is demonstrated by exactly mapping the 1D site-decorated $q$-state Potts model in an external magnetic field onto a zero-field bond-decorated $J_1$-$J_2$ $q$-state Potts model with conventional geometric frustration (Appendix~\ref{Appendix:mapping}).
These results establish a versatile framework for engineering controlled fast state-flipping switches in 1D systems. 

\ignore{
It is known that exact solutions of 1D Potts models can be formulated in terms of the transfer matrix $\mathbb{T}$~\cite{Kramers_Wannier_PR_41_transferMatrix,Glumac_JPA_93_Potts_LR}.  The challenge in finding analytic results for a complete, intuitive understanding of underlying physics arises from the arbitrarily large $q$, e.g., thinking about diagonalization of a $100^{100^{100}}\times 100^{100^{100}}$ matrix; at a glance, this seems intricate. However, for the simplest 1D Potts model with nearest-neighbor interaction $J$ in an external magnetic field $h$, the order of $\mathbb{T}$ is $q$ but can be reduced to $2$~\cite{Glumac_JPA_94_Potts_J1-h} due to the symmetry group $S_q$. Frustrated Potts models are much more difficult because the order of $\mathbb{T}$ is $q^L$ where $L\ge2$ is the number of sites needed to construct $\mathbb{T}$. Previous studies remarkably reduced the task to numerical calculations for an effective $q^{L-1}\times q^{L-1}$ matrix in the integer-$q$ formalism of the transfer matrix---and for an effective $2\times 2$ matrix in the continuous-$q$ formalism where physics is less transparent---however, short of analytic exact results~\cite{Glumac_JPA_93_Potts_LR}. Given the traditional knowledge that the 1D models (with short-range interactions unless specifically noted) is less intriguing due to lacking finite-temperature phase transition and the apparent difficulty due to arbitrary $q$, no surprise that not much efforts had been devoted to analytically solve 1D frustrated Potts models.}

In addition, 1D $q$-state Potts models with finite-temperature UNPC provide a rigorous test ground for AI in science, since solving the transfer matrix $\mathbb{T}$ for large $q$ (e.g., $q=10^{{{10}^{10}}}$), identifying UNPCs and their underlying mechanisms, and extracting meaningful insights from such an expansive parameter space pose daunting challenges, not to mention exact analytic solutions in closed form. It was found in the 1000-Scientist AI Jam Session that OpenAI's reasoning model \texttt{o3-mini-high}, a general-purpose large language model (LLM), could reproduce a manuscript solving a newly proposed site-decorated Ising model ($q=2$) by doing its own math; e.g., the AI-derived equation determining the crossover temperature $T_0$ is astonishingly more elegant~\cite{Yin_site}. Next, \resub{the need to explore UNPC in the Potts model motivated our use of AI to solve the half-century-old open problem in statistical mechanics}---the 1D $J_1$-$J_2$ Potts model for general $q$. The AI's exact analytic solution for $q=3$ bootstrapped the human scientist to generalize it to the full solution for arbitrary $q$ and
obtain an intuitive understanding of the rich phase behavior in this fundamental frustrated model~\cite{Yin_Potts_J1-J2_1D}. 

These developments have fueled the current topical questions as to whether and how AI could lead a project from concept to execution\red{---particularly how well a frontier foundation LLM could perform in exploring science frontiers based on retrieval-augmented generation with latest research results, as it takes time---uncontrollable by the users---for the LLM to be fine-tuned with those results.} 
Here, \texttt{o3-mini-high} was asked to read the aforementioned two manuscripts \red{before their publications}~\cite{Yin_Potts_J1-J2_1D,Yin_site} 
and suggest future work. The AI pointed out five directions with the first one being the present study. Then, it integrated the models and algorithms in Refs.~\cite{Yin_Potts_J1-J2_1D,Yin_site} to carry out all the setup, derivation, coding, and drafting the math-heavy technical sections \red{(i.e., the equations and their descriptions in Secs.~\ref{section:site_model} and \ref{section:bond}, while the rest was written by the author)} in a few minutes with minimal human interference---most insightfully it predicted exactly how to identify UNPCs and the half ice, half-fire mechanism for arbitrary $q$---thus earning its meritorious AI-co-led status (see Appendix~\ref{Appendix:AI} for details). 

\ignore{The rest of the paper is organized as follows: We present the studies of the standard 1D $q$-state Potts models with minimal site decoration (Fig.~\ref{Fig:structure}a) and bond decoration (Fig.~\ref{Fig:structure}b) in Sections II and III, respectively. The results for the two kinds of decorated models are similar in the parameter space of interest---with one significant difference: As the interaction between the backbone spins increases, $T_0$ remains unchanged in the site-decorated model (Type-I UNPC) but shifts toward a finite temperature in the bond-decorated Potts model with $q>2$, giving rise to Type-II UNPC.}

\section{Site-decorated model\label{section:site}}

\subsection{Model, mapping, and solution\label{section:site_model}}

The Hamiltonian of the site-decorated Potts model correctly generated by the AI based on Ref.~\cite{Yin_site} is given by
\begin{eqnarray}    
\label{eq:H}
H &=& H_a+H_b,\nonumber\\
H_a &=& -J\sum_{i=1}^{N}\delta(\sigma_i,\sigma_{i+1}) -  h\mu_a\sum_{i=1}^{N}\delta(\sigma_i,1),\nonumber\\
H_b &=& -J_{ab}\sum_{i=1}^{N}\delta(\sigma_i,b_i) -h\mu_b\sum_{i=1}^{N}\delta(b_i,1),
\end{eqnarray}
where $H_a$ describes the backbone of the single chain with $\sigma_i \in \{1,2,3,...,q\}$ being the spin variable at site $i$ (green balls, referred to as \emph{type-a} spins) and $J>0$ the ferromagnetic interaction (red bonds). $\delta({\sigma_{i},\sigma_{i+1}})$ is the Kronecker delta (which equals 1 if  $\sigma_{i}=\sigma_{i+1}$ and 0 otherwise). $H_b$ describes the decorated parts, where $b_i\in \{1,2,3,...,q\}$ (gray balls, referred to as \emph{type-b} spins) couples to type-\emph{a} spins with the antiferromagnetic interaction $J_{ab}<0$. $h$ depicts the magnetic field, and $\mu_a$ and $\mu_b$ the magnetic moments of type-\emph{a} and type-\emph{b} spins, respectively. The relationship of $\mu_b > \mu_a > 0$ is used to represent ferrimagnetism. We define $h_a\equiv h\mu_a$ and $h_b\equiv h \mu_b$ for shorthand notation. $N$ is the total number of unit cells and $\sigma_{N+1}\equiv\sigma_{1}$, $b_{N+1}\equiv b_{1}$, namely, the periodic boundary condition.

In the thermodynamic limit $N\to\infty$, the partition function $Z=\mathrm{Tr}\,e^{-\beta H}=\lambda^{N}$, where $\lambda$ is the largest eigenvalue of the transfer matrix $\mathbb{T}$ and $\beta=1/(k_\mathrm{B}T)$ with $T$ being the absolute temperature and $k_\mathrm{B}$ the Boltzmann constant.  The free energy per unit cell is given by
\begin{equation}
    f=\lim_{N\to\infty}-\frac{1}{N\beta} \ln Z=-\frac{1}{\beta} \ln\lambda.
\end{equation} 
$f$ determines physical properties such as the entropy per unit cell $S=-\partial f/\partial T$, the specific heat $C_v=T\partial S/\partial T$, the backbone-spin magnetization $m_a=\langle \delta(\sigma_i,1) \rangle=-\partial f/\partial h_a$, the decorating-spin magnetization $m_b=\langle \delta(b_i,1) \rangle=-\partial f/\partial h_b$, etc. 

A key signature of UNPC is the abrupt change in the backbone magnetization. The crossover temperature \(T_0\) is determined by the condition that \(m_a\) attains its symmetric, fully disordered value, i.e.,
\begin{equation}
m_a(T_0) = \frac{1}{q}.
\label{eq:T0}
\end{equation}
Figures~\ref{Fig:structure}(c) and \ref{Fig:structure}(d) show that a UNPC does not exist in the standard (undecorated) Potts models.

To directly demonstrate that a UNPC is driven by a hidden state where one sublattice is ``ice'' (frozen, ordered) and the other is ``fire'' (fully disordered), we calculate the total entropy and identify any abrupt jumps at $T_0$ (or rapid changes over an ultranarrow $T$ range) that indicate a latent-heat analog even though the transition is a crossover. An entropy jump of \(k_\mathrm{B}\ln q\) per unit cell, together with $m_a$ jumping up to the fully ordered value $1$ and $m_b$ jumping down to the fully disordered value $1/q$, provides compelling evidence for the half-ice ({\typea} spins), half-fire ({\typeb} spins) mechanism.

The first step for calculating $Z$ is to construct the transfer matrix. Since the system has two spins per unit cell (one type $a$ and one type $b$), the order of the transfer matrix is $q^2$. After reading Refs.~\cite{Yin_site,Yin_Potts_J1-J2_1D}, the AI first attempted to apply the maximally symmetric subspace (MSS) method~\cite{Yin_Potts_J1-J2_1D} to reduce the $q^2\times q^2$ matrix to $2\times 2$ but quickly realized that this can be easily done only for $h=0$. Then, the AI was reminded that the decorating spins can be summed out exactly~\cite{Yin_site}. The AI immediately reacted to first use this algorithm to reduce the transfer matrix to $q\times q$ and then used the MSS method to further reduce it to $2\times 2$. 

The symmetric $q\times q$ transfer matrix after summing over the decorating spins is given by
\begin{equation}
\label{eq:Teff}
\mathbb{T}_{\sigma\sigma'}= e^{\beta J\,\delta(\sigma,\sigma')+\frac{\beta h_a}{2}[\delta(\sigma,1)+\delta(\sigma',1)]}\,\sqrt{A(\sigma)A(\sigma')},
\end{equation}
where $A(\sigma)$ the on-site weight for a fixed backbone spin $\sigma$ is given by
\begin{equation}
\label{eq:A}
A(\sigma)=\sum_{b=1}^q e^{\beta J_{ab}\,\delta(\sigma,b)+\beta h_b\,\delta(b,1)}.
\end{equation}
Explicitly, one finds
\begin{eqnarray}
&A(1)&\;= e^{\beta(J_{ab}+h_b)}+(q-1),\nonumber\\
&A(\sigma\neq 1)&\;= e^{\beta J_{ab}}+e^{\beta h_b}+(q-2). \label{eq:A}  
\end{eqnarray}
\ignore{the partition function becomes
\begin{equation}
Z=\sum_{\{\sigma_i\}}\prod_{i=1}^{N}\left\{ e^{\beta J\,\delta(\sigma_i,\sigma_{i+1})+\beta h_a\,\delta(\sigma_i,1)} A(\sigma_i)\right\}.
\end{equation}
}
$\mathbb{T}_{\sigma\sigma'}$ can be cast in the standard Potts form
\begin{equation}
\label{eq:PottsTM}
\mathbb{T}^{\rm Potts}_{\sigma\sigma'} = C \, e^{\beta J\,\delta(\sigma,\sigma')+\frac{\beta h_{\rm eff}}{2}[\delta(\sigma,1)+\delta(\sigma',1)]},
\end{equation}
with the overall constant $C = A(\sigma\neq 1)$,
and the \emph{effective magnetic field} defined as
\begin{equation}
\label{eq:heff}
h_{\rm eff}= h_a+\frac{1}{\beta}\ln\!\left[\frac{A(1)}{A(\sigma\neq1)}\right].
\end{equation}

Since $\mathbb{T}^{\rm Potts}$ is invariant under permutations of the $q-1$ states $\{2,3,\dots,q\}$, we introduce two maximally symmetric $q\times 1$ basis vectors:
\begin{eqnarray}
    |\phi_1\rangle &=&|1\rangle=(1,0,0,0,\dots,0)^\top,\nonumber\\
    |\phi_2\rangle &=& \frac{1}{\sqrt{q-1}}\sum_{s=2}^{q}|s\rangle=\frac{(0,1,1,1,\dots,1)^\top}{\sqrt{q-1}}.
    \label{eq:MSS}
\end{eqnarray}
Projecting the \(q\times q\) matrix (\ref{eq:PottsTM}) onto this subspace, we obtain a reduced \(2\times2\) matrix \(\mathbb{T}_2=\begin{pmatrix} u & w \\ w & v\end{pmatrix}\), with elements being
\begin{eqnarray}
\label{eq:site_uvw}
u &=& \langle\phi_1|\mathbb{T}^{\rm Potts}|\phi_1\rangle = C\, e^{\beta J+\beta h_{\rm eff}},\nonumber\\ 
v &=& \langle\phi_2|\mathbb{T}^{\rm Potts}|\phi_2\rangle = C\,[e^{\beta J}+(q-2)],\\ 
 w&=& \langle\phi_1|\mathbb{T}^{\rm Potts}|\phi_2\rangle = C\,\sqrt{q-1}\;e^{\beta h_{\rm eff}/2}.\nonumber
\end{eqnarray}

The largest eigenvalue of $\mathbb{T}$ is the larger eigenvalue of $\mathbb{T}_{2}$ given by
\begin{equation}
\lambda=\frac{u+v}{2}+\sqrt{\left(\frac{u-v}{2}\right)^2+w^2},
\label{eq:lambda}
\end{equation}
whose corresponding normalized eigenvector is given by
$
\mathbf{v} = (c_1,c_2)^\top,
$
with $c_1^2 + c_2^2 = 1$ and \(X = \frac{c_1}{c_2}=\frac{\lambda-v}{w}\). 

In the original Potts basis, the state \(|1\rangle\) is represented by \(|\phi_1\rangle\), so the backbone magnetization is
\begin{equation}
m_a = \bigl\langle \delta(\sigma,1)\bigr\rangle
    = |\langle \phi_1|\mathbf{v}\rangle|^2
    = c_1^2 
=\frac{X^2}{1+X^2}.  
\label{eq:ma}
\end{equation}

The decorating-spin magnetization is given by
\begin{equation}
m_b = m_a\,P_1 + (1-m_a)\,P_2,
\label{eq:mb}
\end{equation}
where the conditional probabilities are given by
\begin{eqnarray}
 P_1 &= P(b=1\,|\,\sigma=1)=\frac{e^{\beta(J_{ab}+h_b)}}{e^{\beta(J_{ab}+h_b)}+(q-1)},\nonumber\\
P_2 &= P(b=1\,|\,\sigma\neq1)=\frac{e^{\beta h_b}}{e^{\beta J_{ab}}+e^{\beta h_b}+(q-2)}.  
\end{eqnarray}
Note that the half-ice, half-fire state features $m_a=1$ and hence $m_b=1/q$. This can be realized by $P_1=1/q$ at $h=(-J_{ab})/\mu_b$ for arbitrary $q$, defining a characteristic field.

Particularly for the site-decorated model, the criterion for finding the crossover temperature \(T_0\), Eq.~(\ref{eq:T0}), is equivalent to the vanishing of the effective field at $T_0$,  $h_{\rm eff}=0$:
\begin{equation}
h_a + \frac{1}{\beta_0}\ln\!\left[\frac{e^{\beta_0(J_{ab}+h_b)}+(q-1)}{e^{\beta_0 J_{ab}}+e^{\beta_0h_b}+(q-2)}\right] = 0,
\label{eq:T0heff}
\end{equation}
with \(\beta_0 = 1/(k_BT_0)\). Hence, $T_0$ is independent of $J$; the crossover width $\propto e^{-\beta_0 J}$ can be narrowed exponentially at fixed $T_0$ by increasing $J$. In the limit $h\to0$,
\begin{equation}
T_0(h\to 0)=\frac{J_{ab}}{k_\mathrm{B}}
\left[\ln\!\frac{\mu_b-(q-1)\mu_a}{\mu_b+\mu_a}\right]^{-1}
\label{eq:T0h0}
\end{equation}
\resub{has a finite-temperature solution} for $\mu_b > (q-1)\mu_a$, \resub{given $\mu_b>\mu_a>0$ and $J_{ab}<0$. Note that $T_0(h\to 0)>0$ also holds for $J_{ab}>0$, $\mu_a\mu_b<0$, and $|\mu_b|>|\mu_a|$; such a parameter space could be realized in mixed $3d$--$5d$ compounds~\cite{Yin_g,Yin_PRL_Sr3CuIrO6}.}

\begin{figure}[t]
\includegraphics[width=\columnwidth]{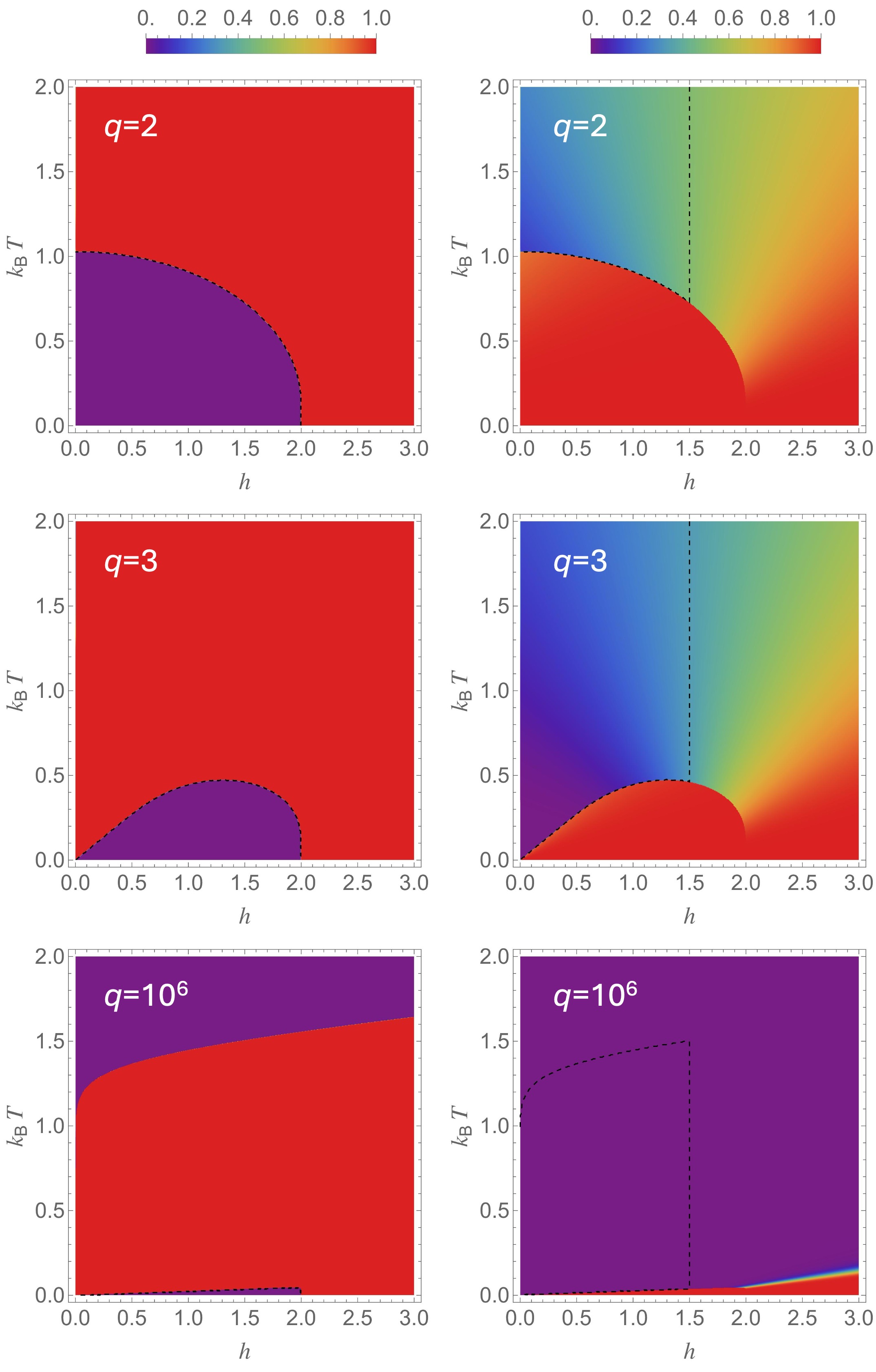}
\caption{Density plots of the backbone magnetization $m_a$ (left panels) and the decorated-spin magnetization $m_b$ (right panels) as functions of the magnetic field $h$ and temperature $T$ for three typical $q$ values: $2, 3, 10^6$. The dashed lines indicate $m_a=1/q$ or $m_b=1/q$, i.e., the fully disordered ``on fire'' state. Here, $\mu_b=4/3$, $\mu_a=1$,  $J_{ab}=-2$, and $J=20$. \resub{The color scale maps $m_a$ or $m_b$ from 0 (purple) to 1~(red), as indicated.}}
\label{fig:magnetizations}
\end{figure}

\begin{figure}[t]
\includegraphics[width=0.95\columnwidth]{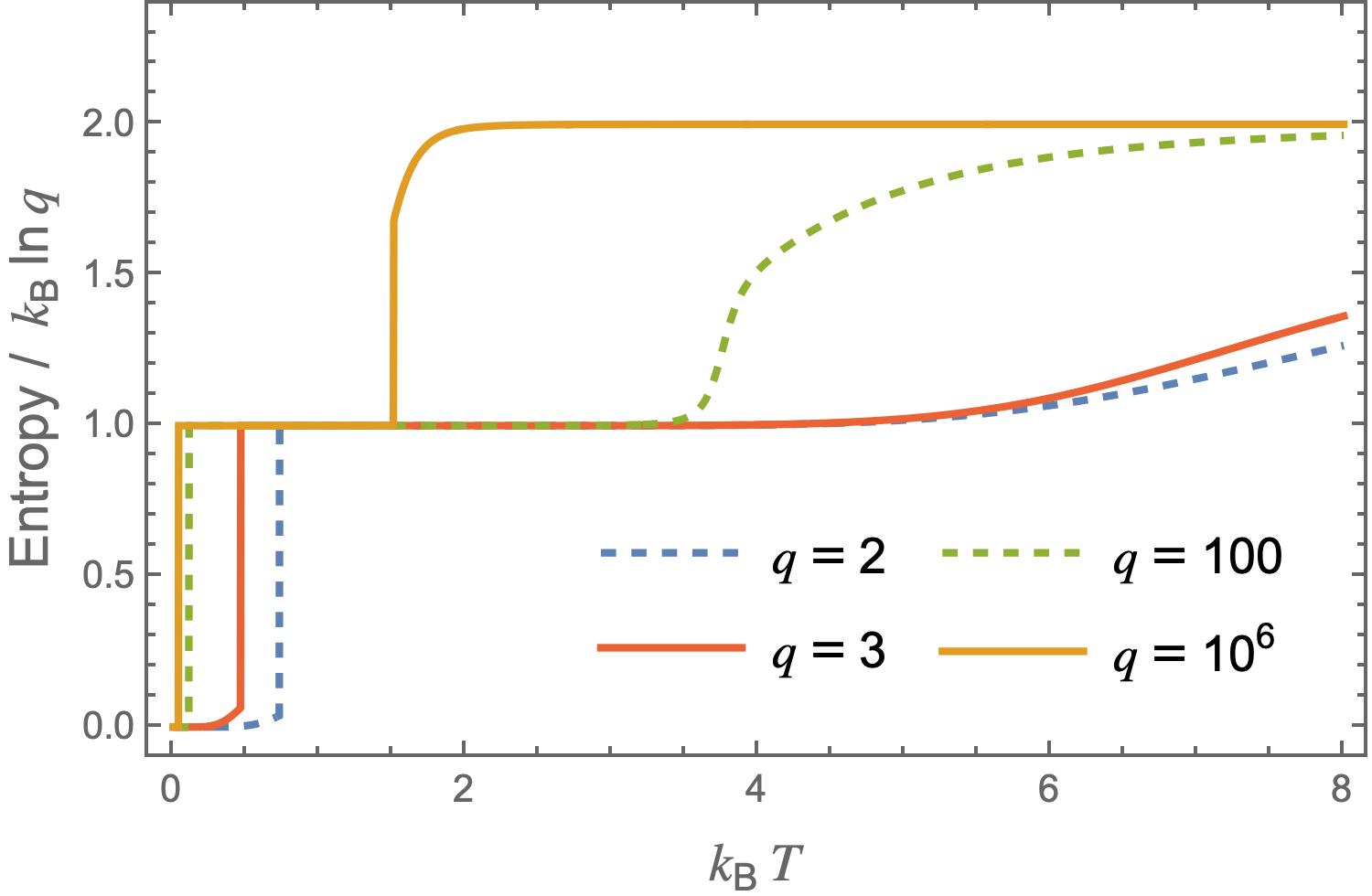}
\caption{Temperature dependence of the normalized entropy per unit cell $S/(k_\mathrm{B}\ln q)$ at $h=(-J_{ab})/\mu_b$ for four typical $q$ values: $2, 3, 100, 10^6$. The jump from 0 to 1 indicates a half-ice, half-fire UNPC. For $q=10^6$, the jump from 1 to 2 indicates a high-$T$ UNPC. Here, $\mu_b=4/3$, $\mu_a=1$,  $J_{ab}=-2$, and $J=20$.}
\label{fig:entropy}
\end{figure}

\subsection{Results and discussion}

Figure~\ref{fig:magnetizations} shows density plots for $m_a$ (left panels) and $m_b$ (right panels) in the $h$-$T$ plane for three typical $q$ values: $2, 3, 10^6$ with $\mu_b=4/3$, $\mu_a=1$, $J_{ab}=-2$, and $J=20$. Compared with Figs.~\ref{Fig:structure}(c) and \ref{Fig:structure}(d) for the undecorated models, the most salient feature in Fig.~\ref{fig:magnetizations} (left panels) is the emergence of the UNPC at $T_0$ (the dashed lines meaning $m_a=1/q$) between two phases with $m_a=0$ (purple $T<T_0$ region) and $m_a=1$ (red $T>T_0$ region), respectively, with the $T_0$ line extending from $h=0$ to $h=(-J_{ab})/\mu_a$ for all $q$. 
In the ground state, since $\mu_b > \mu_a$, the {\typeb} spins are fully ordered in state 1 by the field, which forces the {\typea} spins to order ($J>0$) or disorder (for $J=0$) in one of the $q-1$ states other than state 1 due to the antiferromagnetic coupling $J_{ab}<0$. In other words, the energy cost is $-J_{ab}$ per unit cell for the {\typea} spins to order in state 1, which can be overcome by the field larger than $h_c=(-J_{ab})/\mu_a$, giving rise to the zero-temperature critical point at $h=h_c$ and for $J=0$ the half-fire, half-ice critical state. Likewise, given an excited state where the {\typea} spins are fully ordered in state 1 (``frozen''), the {\typeb} spins would be fully disordered (``on fire'') at the field $h_f=(-J_{ab})/\mu_b<h_c$. Thus,
the half-ice, half-fire mechanism is most easily seen from the emergence of the $m_b=1/q$ dashed lines (right panels) at $h=h_f$ in the region corresponding to the $m_a=1$ phase. 

Further, the entropy jump per unit cell, $\Delta S$, from zero to $k_\mathrm{B}\ln q$ at $T_0$ (Fig.~\ref{fig:entropy}) manifests that $m_b=1/q$ at $h=h_f$ is not only the average magnetization of the {\typeb} spins but an indication that every {\typeb} spin is fully disordered. Meanwhile, $m_a=1$ means that every {\typea} spin is fully ordered.   

\ignore{
The UNPC occurs for $h<h_c$ with the free energy per unit cell being $f(T<T_0)\simeq-h\mu_b$ and $f(T>T_0)\simeq-h\mu_a-h\mu_b/q-J_{ab}/q - T \Delta S$. $T_0(h)$ can be estimated by the level crossing 
$f(T<T_0)=f(T>T_0)$,
\begin{equation}
    T^\mathrm{est}_0(h)\simeq \frac{(h\mu_b-h\mu_a)-(h\mu_b+J_{ab})/q}{\Delta S}.
\end{equation}
In particular,
\begin{equation}
  T_0(h=h_f)\simeq \frac{h_f(\mu_b-\mu_a)}{k_\mathrm{B}\ln q},  
\end{equation}
which decreases as $q$ increases. 
}

\begin{figure}[t]
\includegraphics[width=\columnwidth]{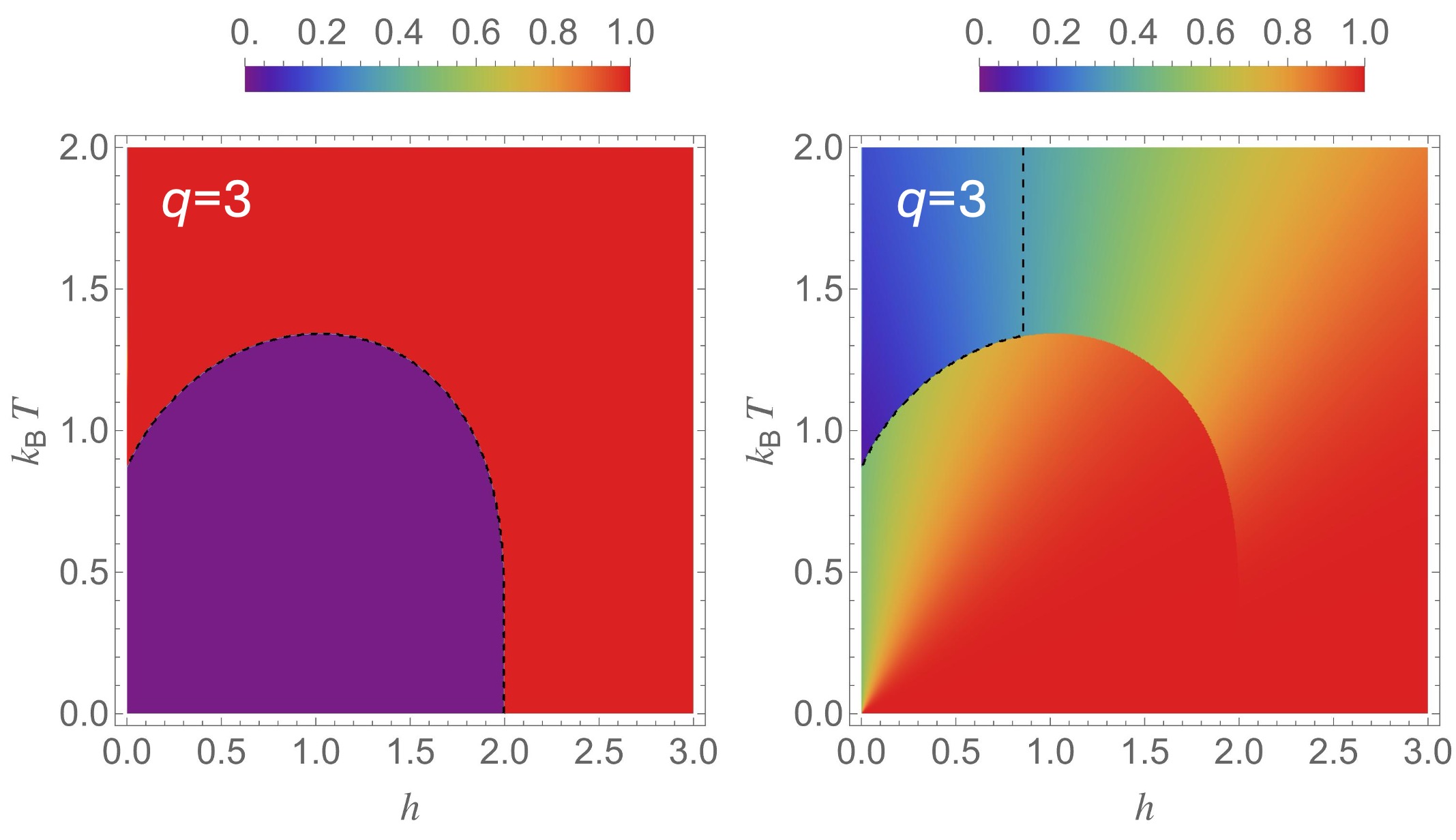}
\caption{Density plots of $m_a$ (left panels) and $m_b$ (right panels) in the $h$-$T$ plane for $q=3$ in the 1D site-decorated Potts model. The dashed lines indicate $m_a=1/q$ or $m_b=1/q$, i.e., the fully disordered ``on fire'' state. Here, $\mu_b=7/3$, $\mu_a=1$,  $J_{ab}=-2$, and $J=20$. \resub{The color scale maps $m_a$ or $m_b$ from 0 (purple) to 1 (red), as indicated.}}
\label{fig:q3_mub2}
\end{figure}

Now, we focus on the differences from the Ising model ($q=2$). The salient feature for $q>2$ is the emergence of a $T_0$ dome as shown in Fig.~\ref{fig:magnetizations}, namely, the maximum of $T_0$ as a function of $h$ appears in between $h=0$ and $h=h_c$. For $q=2$, Eq.~(\ref{eq:T0heff}) is simplified to~\cite{Yin_site} 
\begin{equation}
\tanh\left(\frac{-\beta_0 J_{ab}}{2}\right)
=\frac{\tanh\left(\tfrac{\beta_0 h\mu_a}{2}\right)}{\,\tanh\left(\tfrac{\beta_0 h\mu_b}{2}\right)}.   
\label{eq:T0q2}
\end{equation}
Therefore, $T_0(h)$ is strictly decreasing for $h>0$ and its global maximum occurs in the limit $h\to 0$ for $q=2$. 
However, for $q>2$, Eq.~(\ref{eq:T0h0}) determines that $T_0(0^+)=0$ for $1<\mu_b/\mu_a < q-1$, 
as shown in Fig.~\ref{fig:magnetizations}, where $\mu_b/\mu_a=4/3$. Moreover, even for $\mu_b/\mu_a > q-1$ and hence $T_0(0^+)>0$, the $T_0$ dome persists for $q>2$, as shown in Fig.~\ref{fig:q3_mub2} for $q=3$, where $\mu_b/\mu_a=7/3$. This conclusion is general; e.g., we found $T_0(h_f) > T_0 (0^+)$ for $q>2$ and $\mu_b>\mu_a$, as shown in Fig.~\ref{fig:T0} (Appendix~\ref{Appendix:T0dome}). 

In addition, Figs.~\ref{fig:magnetizations} and \ref{fig:entropy} also reveal that for large $q$, there is a high-temperature UNPC with an entropy jump from $k_\mathrm{B}\ln q$ to $2k_\mathrm{B}\ln q$, i.e., the fully disordered paramagnetic phase, in sharp contrast to the usual broad crossover to the paramagnetic phase as seen for $q=2, 3, 100$ in  Fig.~\ref{fig:entropy}. 

The emergence of the low-$T$ and high-$T$ UNPC can be intuitively understood in the following unified picture thanks to the analytic form of $\mathbb{T}_2$, Eq.~(\ref{eq:site_uvw}): Both the diagonal elements, $u$ and $v$, contain items proportional to $e^{\beta J}$, while the off-diagonal element $w$ does not. Thus, sufficiently large $\beta_0 J$ produces a UNPC with the crossover width $\propto e^{-\beta_0 J}$. The low-$T$ UNPC results from a level crossing from $v$ to $u$ within this ultranarrow temperature range at $T_0$ as $h_\mathrm{eff}$ changes sign. For $q$ large enough to outperform $e^{\beta J}$, the level crossing switches back to $v$ because $u\propto q$ and $v\propto q^2$, resulting in the high-$T$ UNPC.

The fact that the half-ice, half-fire-driven UNPC takes place at $h_\mathrm{eff}=0$ makes the ice-fire mechanism exactly applicable to higher-dimensional site-decorated Potts models, whose exact solution is known for $h=0$ but generally unknown for nonzero $h$. An extremely sensitive response of the spin states to a slight change in the $h$ or $T$ values along the $T_0(h)$ curve is thus expected, like the $q=2$ Ising case~\cite{Yin_site}. 

\section{Bond-Decorated Model\label{section:bond}}

Next, \texttt{o3-mini-high} was asked to read
Ref.~\cite{Yin_Ising_III_PRL}. As shown in Fig.~\ref{Fig:structure}(b), a minimal 1D bond-decorated Potts model generated by the AI based on Ref.~\cite{Yin_Ising_III_PRL} is described by
\begin{eqnarray}    
\label{eq:H}
H &=& H_a+H_b, \nonumber\\ 
H_a &=& -J\sum_{i=1}^{N}\delta(\sigma_i,\sigma_{i+1}) -  h\mu_a\sum_{i=1}^{N}\delta(\sigma_i,1),\\
H_b &=& -J_{ab}\sum_{i=1}^{N}[\delta(\sigma_i,b_i)+\delta(b_i,\sigma_{i+1})] -h \mu_b\sum_{i=1}^{N}\delta(b_i,1).\nonumber
\end{eqnarray}

The AI generated the minimal 1D bond-decorated Potts model corresponding to Fig.~\ref{Fig:structure}(b). Then, it automatically summed out the decorated parts to reduce the transfer matrix to $q\times q$---learned from its mistake in handling the site-decorated model---and then use the MSS method to further reduce it to $2\times 2$.

To sum out the decorating spins, we define the bond weight
\begin{equation}
W(\sigma,\sigma') = \sum_{b=1}^q e^{ \beta J_{ab} [\delta(\sigma,b)+\delta(b,\sigma')]
+\beta h_b\,\delta(b,1)}.
\end{equation}
Specifically,
\begin{eqnarray}
    W(1,1) &=& e^{\beta (2J_{ab} + h_b)} + (q-1), \nonumber\\
    W(1,a) &=& W(a,1)=e^{\beta (J_{ab} + h_b)} + e^{\beta J_{ab}} + (q-2), \nonumber\\
    W(a,a) &=& e^{2\beta J_{ab}} + e^{\beta h_b} + (q-2),\nonumber \\
    W(a,b) &=& 2e^{\beta J_{ab}} + e^{\beta h_b} + (q-3), 
\end{eqnarray}
where $a\ne1$, $b\ne1$, $a\ne b$, and $W(a,b)$ is not needed for $q=2$.

\begin{figure}[t]
\includegraphics[width=\columnwidth]{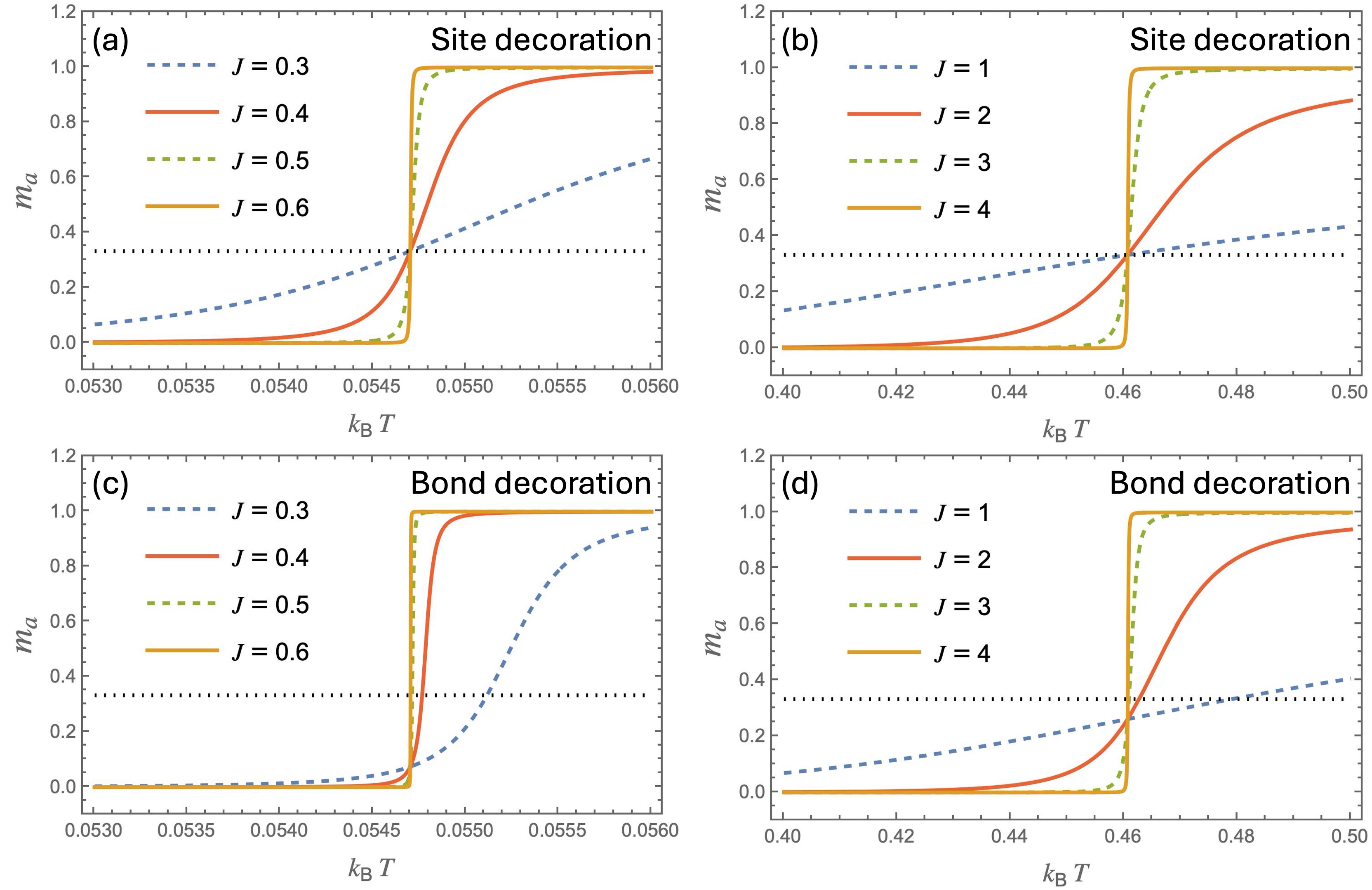}
\caption{Temperature dependence of the backbone magnetization $m_a$ in the $q=3$ Potts model with (a) site decoration at $h=0.1$, (b) site decoration at $h=1.5$, (c) bond decoration at $h=0.1$, and (d) bond decoration at $h=1.5$. Here, $\mu_b=4/3$, $\mu_a=1$, and $J_{ab}=-2$ and $-1$ for site and bond decorations, respectively. $T_0$, at which $m_a=1/q$ (dotted horizontal line), is independent of $J$ for site decoration [panels (a) and (b)]. In contrast, $T_0$ depends on $J$ for bond decoration [panels (c) and (d)] and shifts within an ultranarrow temperature window $\propto e^{-\beta_0 J}$ for an UNPC.}
\label{fig:S1}
\end{figure}

Combining the backbone Boltzmann weight with the bond weight, the \(q\times q\) transfer matrix for the backbone spins is defined by
\begin{equation}
\mathbb{T}_{\sigma\sigma'} = e^{\beta J\,\delta(\sigma,\sigma')
+\frac{\beta h_a}{2}[\delta(\sigma,1)+\delta(\sigma',1)]} \,W(\sigma,\sigma'),
\end{equation}
which cannot be described by an effective magnetic field. \ignore{(The AI recognized this inconvenience and tried to provide approximate solutions. Upon being reminded of exact solution, the AI proceeded.)} However, because \(W(\sigma,\sigma')\) depends only on whether \(\sigma\) and \(\sigma'\) equal 1 or not, the matrix is invariant under any permutation of the \(q-1\) states in \(\{2,\dots,q\}\). We use Eq.~(\ref{eq:MSS}) to project the \(q\times q\) matrix to obtain a reduced \(2\times2\) matrix \(\mathbb{T}_2=\begin{pmatrix} u & w \\ w & v\end{pmatrix}\), with elements
\begin{eqnarray}    
u &=& \langle\phi_1|\mathbb{T}|\phi_1\rangle = e^{\beta J+\beta h_a}\,W(1,1), \nonumber\\
v &=& \langle\phi_2|\mathbb{T}|\phi_2\rangle = e^{\beta J}\,W(a,a)+(q-2)\,W(a,b),\nonumber \\
w &=& \langle\phi_1|\mathbb{T}|\phi_2\rangle = \sqrt{q-1}\,e^{\frac{\beta h_a}{2}}\,W(1,a).
\label{eq:bond_uvw}
\end{eqnarray}
The largest eigenvalue of $\mathbb{T}$ is the larger eigenvalue of $\mathbb{T}_{2}$. 
\ignore{determines the free energy per backbone spin via
\(
f = -\frac{1}{\beta}\ln \lambda,
\)
and the magnetization is obtained from
\(
m_a = -\frac{\partial f}{\partial h_a}=\frac{(\lambda-v)^2}{(\lambda-v)^2+w^2}
\), Equation~(\ref{eq:ma}), 
and \(
m_b = -\frac{\partial f}{\partial h_b}
\).

$T_0$ is determined by $m_a(T_0)=1/q$, Equation~(\ref{eq:T0}). In the limit $h\to0$,
\begin{equation}
\frac{e^{\beta_0 J}(e^{2\beta_0 J_{ab}} - 1) - e^{\beta_0 J_{ab}}- (q-2)}
{e^{\beta_0 J}(e^{2\beta_0 J_{ab}} + q-1) - e^{\beta_0 J_{ab}} - \frac{1}{2}(q-2)}
\;=\;-\frac{\mu_a}{\mu_b}.
\label{eq:bond_T0h0}
\end{equation}
Hence, $T_0$ depends on $J$. However, in the limit $h\to0$ and the large $J$ limit $\beta_0 J\gg q-2$,
\begin{equation}
k_\mathrm{B}T_0=2J_{ab}
\left[\ln\!\frac{\mu_b-(q-1)\mu_a}{\mu_b+\mu_a}\right]^{-1},
\label{eq:T0h0J}
\end{equation}
for $\mu_b > (q-1)\mu_a$ and zero otherwise, which
}

A low-$T$ UNPC results from the level crossing from $v$ to $u$ at $T_0$ with the crossover width $\propto e^{-\beta_0 J}$. Since a {\typeb} spin couples to one and two {\typea} spins in the site- and bond-decorated models, respectively, the results of the two models are similar for sufficiently large $\beta_0 J$ when the energy unit is set as $J_{ab}=-2$ and $J_{ab}=-1$ for the site- and bond-decorated models, respectively. It thus suffices that we focus on discussing the results for the site-decorated model in the main text; similar results for the bond-decorated model are openly available~\cite{data:Wolfram2}.

Both decorations yield the same $J$-independent $T_0$ for $q=2$ (Ising model)~\cite{Yin_Ising_III_PRL,Yin_site}. However, there is a significant difference in the bond-decorated model with $q>2$: As shown in Fig.~\ref{fig:S1}, $T_0$ determined by $m_a(T_0)=1/q$ depends on $J$ and can now shift toward a finite temperature determined in the $J\to\infty$ limit; when UNPC takes place, $T_0$ shifts within an ultranarrow temperature window $\propto e^{-\beta_0 J}$. This provides a new perspective of UNPC, referred to as Type-II UNPC.

\section{Summary}

In summary,  we have exactly solved 1D site- and bond-decorated $q$-state Potts ferrimagnets in an external magnetic field. By summing out the decorating spins and exploiting permutation symmetry, the full \(q^2\times q^2\) transfer matrices reduce to \(2\times2\).  We derive closed‐form expressions for the backbone magnetization \(m_a\), the decorating‐spin magnetization \(m_b\), and the entropy \(S\), and show that an ultranarrow phase crossover  occurs precisely when \(m_a=1/q\).  Entropy jumps of order \(k_B\ln q\) confirm a hidden half-ice, half-fire  mechanism.  For \(q>2\), \(T_0(h)\) exhibits a pronounced dome, and for large \(q\) a secondary high-temperature UNPC emerges. The results for the two kinds of decorated Potts models are similar in the parameter space of interest---with one significant difference: As the interaction between the backbone spins increases, $T_0$ remains unchanged in the site-decorated model (Type-I UNPC) but shifts toward a finite temperature in the bond-decorated Potts model with $q>2$, giving rise to Type-II UNPC. These results generalize the ice-fire-driven ultranarrow phase crossover to the Potts models and provide a platform for engineering fast, entropy-order switches in low-dimensional systems. We demonstrated that a general-purpose LLM can lead the project from concept to execution. 

\section*{acknowledgment}

This work was supported by U.S. Department of Energy (DOE) Office of Basic Energy Sciences (BES) Division of Materials Sciences and Engineering under Contract No. DE-SC0012704.

\section*{DATA AVAILABILITY}
The codes, partly generated by the AI, and data that support the findings of this article are openly available~\cite{data:Wolfram2}.



\appendix
\section{A Framework for AI Contribution Rating}
\label{Appendix:AI}

To measure the AI contribution to research, from minimal \emph{post hoc} verification of human work to fully AI‐led discovery, we introduce the following 
\red{partially ordered nine-category framework} from one dan (1d) to nine dan (9d), named after the skill ranks in the traditional board game Go~\cite{Silver_Nature_16_AlphaGo,Silver_Nature_17_AlphaZero}: 

\begin{enumerate}[label=(\arabic*d)]
  \item \emph{AI-verified}:  
    AI reads the human's complete draft, checks every derivation, and confirms no errors.

  \item \emph{AI-validated}:  
    AI reads the manuscript, rederives key results (and corrects minor issues), and its polished expressions are adopted.

  \item \emph{AI-assisted}:  
    AI performs requested coding, calculations, drafts text snippets, or generates figures under direct human guidance.

  \item \emph{AI-completed}:  
    AI executes all detailed coding, calculations, derivations, and write-up from the human’s outlined goals and framework.

  \item \emph{AI-co-led}:  
    Human and AI independently conceive the project; both human and AI can lead the project from concept to execution.
    
  \item \emph{AI-reshaped}:  
    AI helps pivot the research into a new approach/direction/area that enables the human researcher to make significant progress.

  \item \emph{AI-bootstrapped}:  
    AI provides the crucial seed example to a long-standing problem; the human then recognizes the pattern and completes the general solution.

  \item \emph{AI-driven}:  
    Reversing the course of 7d, AI autonomously generalizes seed examples supplied by the human to the full solution.

  \item \emph{AI-led}:  
    AI conceives the project, completes all work, and the human only reviews and verifies correctness. Example projects include NP problems whose solutions are easy to check but hard to find.     
\end{enumerate}

This system can be divided into two groups: (1) 1d through 5d, in which the human can complete the study without the AI and the AI is employed to accelerate the project, and (2) 6d through 9d, in which the human cannot complete the study without the AI and the AI acts as a scientific discoverer and an inspirational research partner. \red{7d may be considered as an early sign for artificial superintelligence (ASI), while 5d may be an indicator for artificial general intelligence (AGI). The two groups thus represent two distinct tracks toward AGI and ASI, respectively.}


\red{The proposed nine-dan framework should be read as a way to classify how AI contributes to a given piece of research, not as a simple score of how ``good'' or ``intelligent'' a model is. The categories do have an overall direction: 1d is always the least contribution and 9d the greatest, and some subsets are strictly ordered, such as 3d $<$ 4d $<$ 5d and 7d $<$ 8d $<$ 9d. At the same time, the framework is not fully linear. A 2d contribution can sometimes be more important to the scientific outcome than a 3d one, for example, if it produces an unexpectedly decisive improvement. Similarly, 7d need not imply 5d as a subset, because the two emphasize different aspects: 5d is more closely tied to autonomous, general problem solving, whereas 7d may instead reflect superhuman performance on a human-specified task. The assigned category therefore depends not only on the AI model itself, but also on the role it is asked to play in that particular work. Accordingly, a highly capable system may still receive a low-d designation if it is used only in a limited role. For this reason, the framework is best understood as a partially ordered classification of AI contribution.}

Then, the AI mapped its contribution to recent studies of the 1D Ising~\cite{Yin_site}, Heisenberg~\cite{yin_7_CMM_arXiv}, and Potts~\cite{Yin_Potts_J1-J2_1D} models onto AI-validated (2d), AI-reshaped (6d), and AI-bootstrapped (7d), respectively.

As for the present study, the author had derived all the equations before testing the AI without revealing the derivation. Had the human researcher drafted a manuscript---likely in a few months \red{instead of one week with the aid of AI}---and then let the AI read it, the AI's role would have become AI-validated (2d). On the other hand, since the AI predicted independently how to identify the UNPC and the underlying half ice, half-fire mechanism for arbitrary $q$ after reading the solution for $q=2$, had the human researcher not known how to do this, the AI's role would have been ranked as AI-driven (8d). 
This gave us an important lesson that the AI-contribution rating depends on both how much the human researcher knows about solving the problems 
and how much this human knowledge has been passed to the AI. Overall, the AI earned its meritorious status as AI-co-led (5d).

\resub{\emph{Note added.} The disproof of the unit distance conjecture by an internal model at OpenAI, reported in May 2026 and verified by human mathematicians~\cite{9d_Alon_26,9d_Sawin_26}, might be an example of AI-led (9d).
}

\section{Mapping\label{Appendix:mapping}}
\renewcommand{\theequation}{B\arabic{equation}}%

\begin{figure}[t]
\includegraphics[width=0.7\columnwidth]{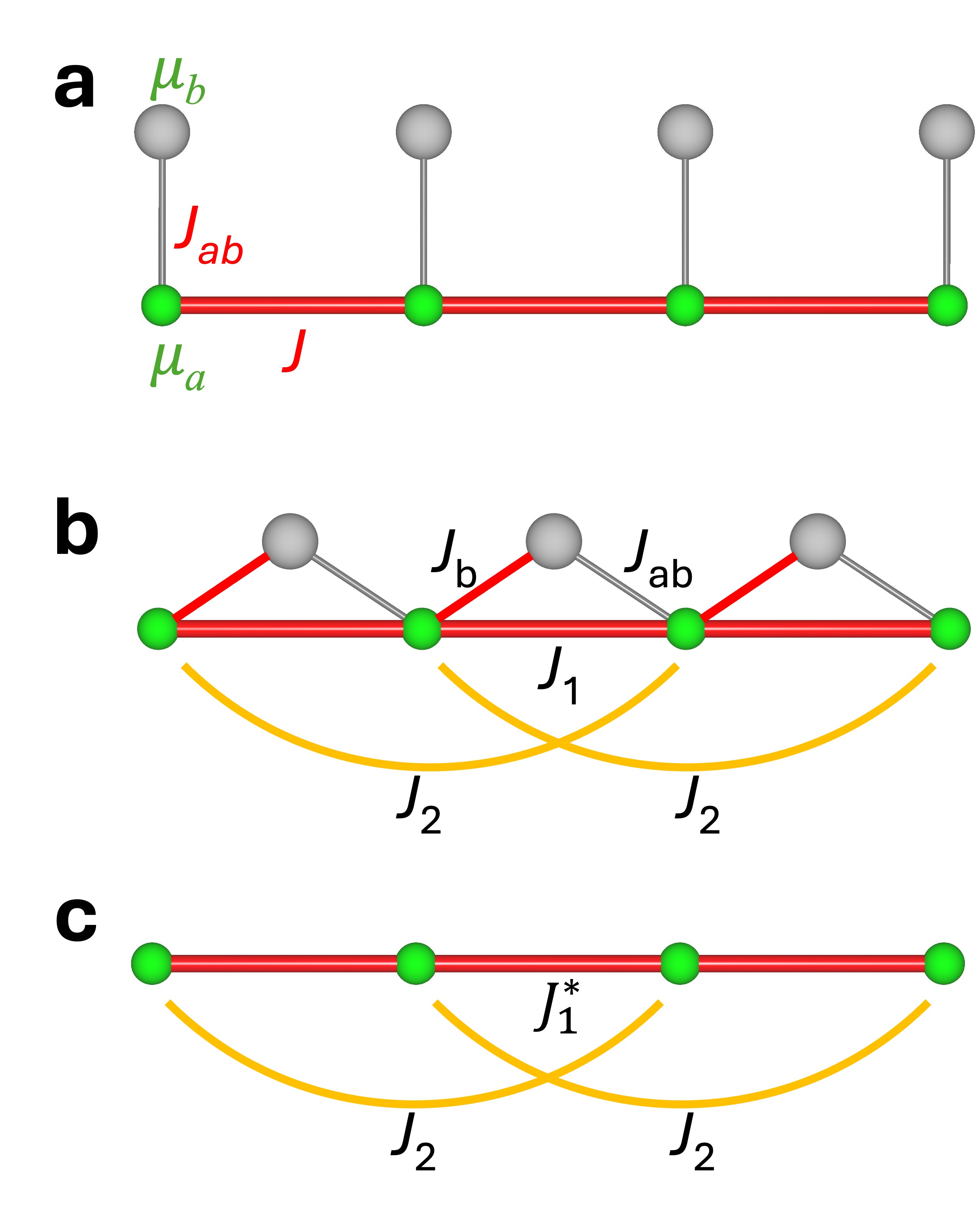}
\caption{Exact mapping among (a) the site-decorated Potts model in the presence of an external magnetic field, (b) a bond-decorated $J_1$-$J_2$ Potts model in the absence of a magnetic field, and (c) an undecorated $J^*_1$-$J_2$ Potts model in the absence of a magnetic field: Model (b) can be mapped exactly onto model (c). Model (a) and model (c) can be mapped exactly onto the same effective Potts model.}
\label{fig:mapping}
\end{figure}

It was recently proven that the 1D site-decorated Ising model ($q=2$) in an external magnetic field [Fig.~\ref{fig:mapping}(a)] can be exactly mapped onto a 1D zero-field bond-decorated $J_1$-$J_2$ Ising model [Fig.~\ref{fig:mapping}(b)], where $J_1$ and $J_2$ are the nearest-neighbor and next-nearest-neighbor interactions between the backbone spins, respectively, and the decorating spins are coupled to two nearest-neighbor backbone spins by interactions $J_b$ and $J_{ab}$, when $J_1=h\mu_a$, $J_b=h\mu_b$, and $J_2=J$~\cite{Yin_site}. 

This mapping holds for general $q$, following a similar proof procedure: 

(1) The bond-decorating spins in Fig.~\ref{fig:mapping}(b) can be summed out exactly for any dimension, resulting in an effective standard $J_1^*$-$J_2$ Potts model [Fig.~\ref{fig:mapping}(c)], with $J_1$ being replaced by a temperature-dependent effective nearest-neighbor interaction~\cite{Potts_RMP_82} 
\begin{equation}
    J_1^*= J_1 + 
\frac{1}{\beta}\ln\!\left[\frac{A(1)}{A(\sigma\neq1)}\right],
\label{eq:Jeff}
\end{equation}
where $A(1)$ and $A(\sigma\neq1)$ are defined in Eq.~(\ref{eq:A}) with the following replacement $h\mu_a\to J_1$, $h\mu_b\to J_b$, and $J_{ab}\to J_{ab}$. With the above replacement, Eq.~(\ref{eq:Jeff}) appears to be the same as Eq.~(\ref{eq:heff}).

(2) The 1D $J_1^*$-$J_2$ Potts model can be exactly mapped onto the 1D Potts model with $J_2$ acting as the nearest-neighbor interaction and $J_1^*$ as an external field~\cite{Yin_Potts_J1-J2_1D}. That is, Eq.~(\ref{eq:Jeff}) becomes the same effective magnetic field as Eq.~(\ref{eq:heff}).

Therefore, for $J_2=J$, $J_1^\mathrm{eff}=h_\mathrm{eff}\mu_a$, and $J_b=h\mu_b$, the 1D zero-field bond-decorated $J_1$-$J_2$ Potts model [Fig.~\ref{fig:mapping}(b)] has been mapped exactly onto the 1D site-decorated Potts model in an external magnetic field [Fig.~\ref{fig:mapping}(a)]. \qedblack

The bond-decorated $J_1$-$J_2$ model exhibits conventional geometric frustration when $J_1 J_b J_{ab}<0$, i.e., when the frustration function for any triangle formed by the three interactions is negative~\cite{toulouse1977frustration}. Correspondingly, the site-decorated model is frustrated for $\mu_a\mu_b J_{ab}<0$. In addition, for a UNPC to occur, $J_1$ must be the weakest among all interactions, particularly when $|J_1|\ll |J_2|$. Such a condition is not easily realized for spin-spin interactions. In contrast, in the site-decorated model it is straightforward to achieve $|h\mu_a|\ll |J|$ with a weak magnetic field $h$. Therefore, the mapping offers the dual advantage of employing conventional geometric frustration to elucidate the origin of frustration and collective behavior, while using the magnetic field as a convenient knob for tuning the parameters.

\section{Emergence of the $T_0$ Dome for $q>2$\label{Appendix:T0dome}}

\begin{figure}[t]
\includegraphics[width=0.95\columnwidth]{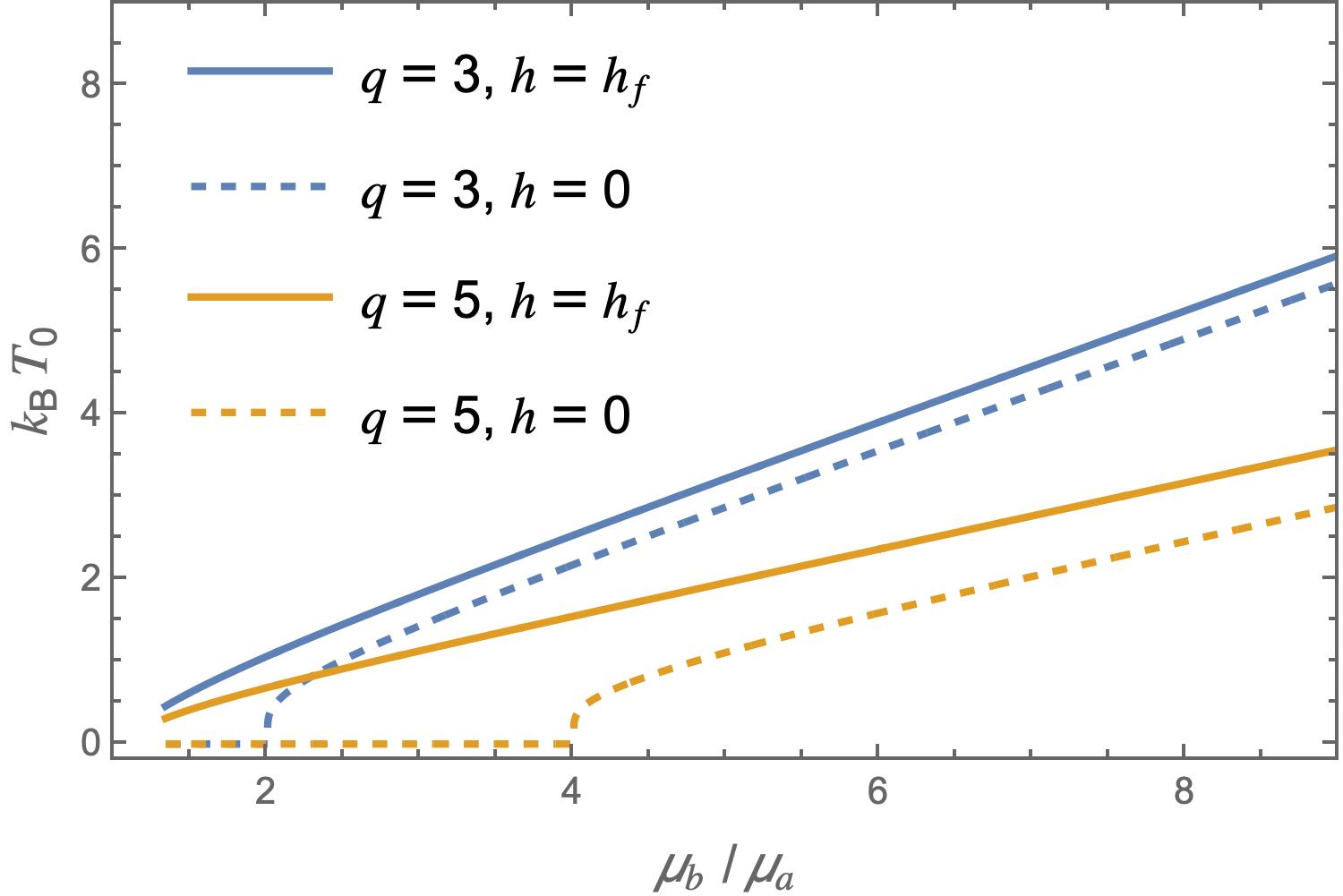}
\caption{$T_0$ at $h_f$ (solid lines) and $T_0$ in the limit $h\to 0$ (dashed lines) as a function of $\mu_b/\mu_a$ for $q=3$ and $q=5$. Here, $\mu_a=1$,  $J_{ab}=-2$, and $J=20$.}
\label{fig:T0}
\end{figure}

Figure~\ref{fig:T0} shows that $T_0(h_f) > T_0 (0^+)$ for $q>2$ and $\mu_b>\mu_a$. On the other end at $h=h_c>h_f$, $T_0(h_c)=0$. Therefore, there exists a $T_0(h)$ dome in between $h=0$ and $h=h_c$. 

\section{Software} 
\texttt{Wolfram Mathematica 14.2} was used to do mathematic derivation and visualize the results. 

\input{Potts_5d_proof.bbl}

\end{document}

%% file: Potts_5d_proof.bbl
%